\documentclass[useAMS]{mn2e}
\usepackage{psfig}
\newif\ifAMStwofonts
\def\be{\begin{eqnarray}}
\def\ee{\end{eqnarray}}
\def\beq{\begin{equation}}
\def\eeq{\end{equation}}

\def\etal{{\it et al.}}

\def\HI{{\hbox{H~$\scriptstyle\rm I\ $}}}
\def\HII{\hbox{H~$\scriptstyle\rm II\ $}}

\def\SiIV{\hbox{Si~$\scriptstyle\rm IV\ $}}
\def\CIV{\hbox{C~$\scriptstyle\rm IV\ $}}
\def\OIII{\hbox{O~$\scriptstyle\rm III\ $}}
\def\nH{{\rm H}}
\def\nHI{{\rm HI}}

\def\lya{Ly${\alpha}\,\,$}

\def\lesssim{\mathrel{\hbox{\rlap{\hbox{\lower4pt\hbox{$\sim$}}}\hbox{$<$}}}}
\def\gtrsim{\mathrel{\hbox{\rlap{\hbox{\lower4pt\hbox{$\sim$}}}\hbox{$>$}}}}

\def\gtsima{$\; \buildrel \over \sim \;$}
\def\ltsima{$\; \buildrel < \over \sim \;$}
\def\prosima{$\; \buildrel \propto \over \sim \;$}
\def\gsim{\lower.5ex\hbox{\gtsima}}
\def\lsim{\lower.5ex\hbox{\ltsima}}
\def\simgt{\lower.5ex\hbox{\gtsima}}
\def\simlt{\lower.5ex\hbox{\ltsima}}

\def\simpr{\lower.5ex\hbox{\prosima}}
\def\la{\lsim}
\def\ga{\gsim}

\def\etal{{\frenchspacing etal. }}
\def\ie{{\frenchspacing\it i.e. }}
\def\eg{{\frenchspacing\it e.g. }}

\def\be{\begin{eqnarray}}
\def\ee{\end{eqnarray}}

\def\CR{{\tt CRASH }}

\title[On the Size of \HII Regions around High Redshift Quasars]
{On the Size of \HII Regions around High Redshift Quasars}

\author[A. Maselli, S. Gallerani, A. Ferrara \&
T.R. Choudhury]{A. Maselli$^{1}$, S. Gallerani$^{2}$, A. Ferrara$^{2}$,
T.R. Choudhury$^{3}$ \\\\
$^1$ Max-Planck-Institut f\"ur Astrophysik, Karl-Schwarzschild-Strasse1,
85748 Garching, Germany\\
$^2$ SISSA/International School for Advanced Studies, via Beirut
2-4, 34014 Trieste, Italy\\
$^3$ Center for Theoretical Studies, Indian Institute of Technology, Karagpur 721302, India}

\date{\today}
\pagerange{\pageref{firstpage}--\pageref{lastpage}} \pubyear{2004}
\begin{document}

\maketitle
\label{firstpage}

\begin{abstract}
We investigate the possibility of constraining the ionization
state of the Intergalactic Medium (IGM) close to the end of
reionization ($z\approx 6$) by measuring the size of the \HII
regions in high-$z$ quasars spectra. We perform a combination of
multiphase smoothed particle hydrodynamics (SPH) and three-dimensional (3D) 
radiative transfer (RT) simulations to reliably
predict the properties of typical high-$z$ quasar \HII regions,
embedded in a partly neutral IGM ($x_\nHI=0.1$). In this work we
assume a fixed configuration for the quasar lifetime and
luminosity, \ie $t_Q=10^7$ yr and $\dot{N_\gamma}=5.2 \times
10^{56}$s$^{-1}$. From the analysis of mock spectra along lines of
sight through the simulated QSO environment we find that the \HII
region size derived from quasar spectra is on average 30 per cent
smaller than the physical one. Additional maximum likelihood
analysis shows that this offset induces an overestimate of the
neutral hydrogen fraction, $x_\nHI$, by a factor $\approx 3$. By
applying the same statistical method to a sample of observed QSOs
our study favors a mostly ionized ($x_\nHI < 0.06$) universe at
$z=6.1$.
\end{abstract}


\section{Introduction}
Studies of the Gunn-Peterson (Gunn \& Peterson 1965, GP) test have now firmly established
that at $z \gtrsim 6$ the mean volume (mass) weighted neutral hydrogen fraction, $x_\nHI$, is higher
than $10^{-3}$ ($10^{-4}$) (e.g. Fan \etal 2006a); a more precise determination is hampered by the strong
sensitivity of \lya photons resonant scattering to even tiny amounts of \HI.
Nevertheless, assessing the exact value of $x_\nHI$ at $z\sim6$  would represent a
crucial information to discriminate between different reionization scenarios and identify
the nature and distribution of ionizing sources.

The size of \HII regions around high-$z$ luminous quasars prior to complete reionization is strongly dependent on
$x_\nHI$. Previous studies (\eg Wyithe \& Loeb
2004; Wyithe \etal 2005) have tried to use this observable to improve upon
the above GP constraints.

In a simple  (pre-overlap) reionization picture in which isolated quasar \HII regions expand into an
intergalactic medium (IGM) with mean neutral hydrogen fraction $x_\nHI$, the (physical) size of the \HII region is
\beq
\label{radius}
 R_d \approx
\left(\frac{3 \dot{N}_\gamma t_Q}{4 \pi n_\nH x_\nHI}\right)^{1/3}
\eeq where $\dot{N}_\gamma$, $t_Q$ are the ionizing photons
emission rate and the quasar lifetime, respectively, and $n_\nH$
is the hydrogen number density. This expression does not account
for the recombinations occurring in the ionized gas, which can be
neglected as typical quasar lifetimes ($t_Q=10^{6-8}$ yr) are much
smaller than the IGM recombination time scales at these epochs.
Also note that eq. \ref{radius} does not hold during the
relativistic phase of the \HII region expansion, which lasts as
long as $\dot{N}_\gamma t \gtrsim (4\pi/3)n_\nH x_\nHI(ct)^3$. For
typical high-$z$ quasars luminosity this phase could last as much
as $t_Q$. However (see \eg White \etal 2003), the size of the {\it
observed} \HII region along the line of sight (LOS) to the quasar
turns out to have exactly the same time evolution as deduced by
assuming an infinite speed of light. In summary,  eq. \ref{radius}
can be safely used to describe the size of the quasar \HII region
observed in quasar absorption spectra and to put constraints on
$x_\nHI$.

In practice, this procedure is affected by a number of possible
sources of error, which range from uncertainties in the quasar
properties (lifetime, luminosity, highly biased environment) to
intrinsic problems in the practical definition of the \HII region
size from the observed spectra. Despite all these problems, the
available set of quasar spectra with a transmission region inside
the GP trough have been exploited to constrain $x_\nHI$. Wyithe \&
Loeb (2004) and Wyithe  \etal (2005) performed a statistical
comparison between observed and predicted radii, and derived a
mean $x_\nHI\gtrsim 0.1$ at $z\approx 6$. An independent study by
Mesinger \& Haiman (2004, hereafter MH04), not involving a direct
measurement of the \HII region size, found $x_\nHI \gtrsim 0.2$ by
modelling observed properties of the \lya and Ly$\beta$ regions of
the spectrum of the $z=6.28$ quasar SDSS J1030+0524.  These
results seem to indicate a largely neutral IGM at those epochs,
somewhat at odd with recent interpretations of the 3-yr WMAP data
(Choudhury \& Ferrara 2006, Gnedin \& Fan 2006). Independent
analysis of the same QSOs spectra by Yu \& Lu (2005) and Fan \etal
(2006a) find much lower values for $x_\nHI$. Furthermore a recent
theoretical analysis by Bolton \& Haenhelt (2006) argues that
current observed spectra are consistent with a very broad range of
$x_\nHI$ values.

In this work we move a step further: by performing a detailed
statistical analysis of mock quasar
spectra extracted from the combined SPH and 3D RT simulations, we quantify the confidence
level at which $x_\nHI$ can be constrained by inverting
eq.\ref{radius}.

\section{Simulations and Results}

We have performed a combination of multiphase SPH and 3D RT
simulations, in order to predict reliably the geometrical shape of the
\HII region around a typical quasar observed at $z\gtrsim6$. High-$z$ luminous
QSOs reside in rare overdense regions where the IGM physical
properties are highly biased (Yu \& Lu 2005). This bias
has been taken into account by using a snapshot centered at $z_Q=6.1$
of the G5 simulation described in Springel \& Hernquist (2003).
With its large computational volume (100$h^{-1}$ comoving Mpc on a side)
and a particle resolution of $2\times 324^3$,
G5 allows us to properly follow the quasar \HII region volume at a sufficiently high
resolution.
The density field is centered on the most massive halo,
$M_{halo} \approx 2.9 \times 10^{12} M_\odot$. This mass is
consistent with that expected for halos hosting high-$z$ luminous quasars (Wyithe
\& Loeb, 2004).

The SPH density field has been mapped on a Cartesian grid with 128$^3$
cells, in order to perform full 3D RT simulations. We used the RT code
\CR (Maselli \etal 2003; Ciardi \etal 2001), which follows the time
evolution of gas ionization state and temperature.
Or spatial resolution does not allow to resolve the IGM clumpiness
at scales above $\approx 0.76$ Mpc comoving;  this prevent us to
correctly account for nonlinear clumps like halos and for their effects on
the overall recombination rate and on shielding.
\begin{figure}
\centerline{\psfig{figure=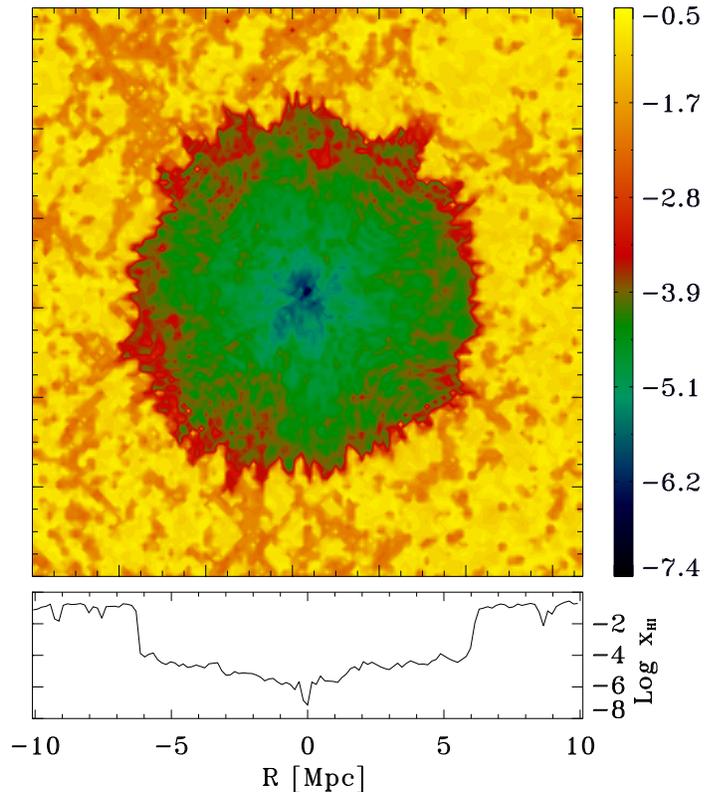,height=10.5cm}}
\caption{Map (upper panel) and cut (lower panel) of the simulated
$x_\nHI$ (logarithmic scale) across the quasar (located at the center
of the box). The quasar \HII region is clearly identified.}
\label{map}
\end{figure}

The quasar is embedded in the most massive halo and we have adopted
a Telfer template (Telfer \etal 2002) in the energy range $13.6 eV - 42 eV$
as its UV spectrum.
For computational economy He physics was not included in the simulations. This might lead to a
underestimate of the ionized gas temperature and could affect the recombination rate.
A higher temperature might increase the inner (resonant) \lya opacity,
essentially due to thermal broadening of the line. In order to quantify this effect, we
have re-calculated some examples of mock spectra along the same l.o.s. used
in the analysis presented in the paper, increasing the temperature by 30 per cent
\footnote{In a previous work
(Maselli etal. 2003) we found that including He typically increases the temperature by 30 per cent
for a mean density of 1 cm$^{-3}$. As in this work the
density are much lower we can safely consider 30 per cent as an
upper limit for the temperature increase due to helium photoheating.}.
Since recombinations are negligible and we have checked that the corresponding thermal broadening
is not affecting the Ly$\alpha$ resonant opacity in a sensible way, He inclusion would not change
the results shown here.

The quasar radiation is sampled by emitting $N_p=10^8$ photon packets.
We have tested numerical convergence by running lower resolution
simulations with 0.5$N_p$ and 0.25$N_p$.
Furthermore we assume $t_Q=10^7$ yr and $\dot{N}_\gamma=5.2\times 10^{56}$
s$^{-1}$; the latter values have been adopted after MH04 analysis of
SDSS J1030+0524. It is fair to note, though, that
the estimates of $t_Q$ and $\dot{N}_\gamma$ should be considered as
tentative because of: (i) the degeneracy among $\dot{N}_\gamma$,
$t_Q$, $x_\nHI$ and the adopted spectral template, (ii) uncertainties
in the quasar redshift, (iii) possible lensing effects, (iv) IGM clumpiness
(see \eg White \etal 2003 for a discussion). However, this
difficulties have little impact on our study as our main aim is to
estimate the confidence level at which $x_\nHI$ can be extracted from obserations.

Initially, the IGM is in photoionization equilibrium with an uniform ionizing background (produced by sources other
than the considered quasar) with a mean photoionization rate $\Gamma_{12}=0.015$ s$^{-1}$, yielding $\langle x_\nHI \rangle =  0.1$. This
value corresponds to the lower limit found by previous works (MH04; Wyithe \& Loeb, 2004); in a
forthcoming paper we will present results relative to other initial $x_\nHI$ values.
In this work we do not account for spatial fluctuations in the ultraviolet background (UVB), 
which are believed to be in place even in a highly ionized
IGM at this epoch (Maselli \& Ferrara 2005).
The main implication of the assumption of a uniform background
is a possible underestimate of the deviation from spherical
symmetry of the physical HII region. Accounting for a possible inhomogeneous
background is expected to increase the dispersion of the physical radius of
the HII region along different lines of sight.

\begin{figure}
\centerline{\psfig{figure=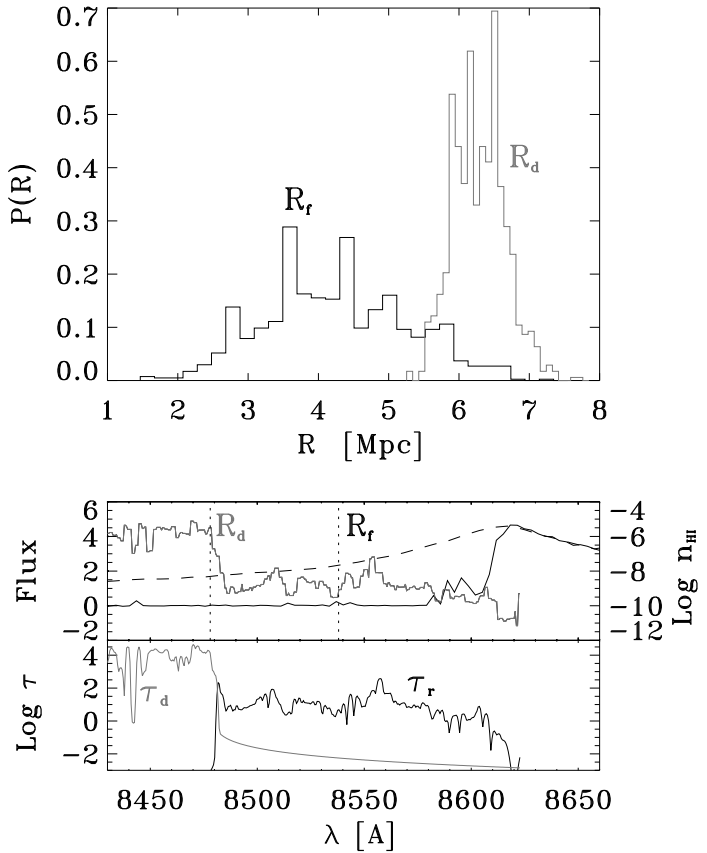,height=11.5cm}}
\caption{{\it Upper panel:} Probability distribution function for $R_d$ and $R_f$ (physical units) using 1000
LOS through the simulation box. The offset between the two
distributions quantifies the apparent shrinking (see text). {\it Central panel:}
Illustrative template (dashed line) and absorbed (solid dark) spectra,
along with the $n_\nHI$ density distribution (light gray) as a function
of observed wavelength, for a representative LOS. {\it Bottom panel:}
Contributions to the total GP optical depth, $\tau$, from neutral hydrogen within ($\tau_r$, dark)
and outside ($\tau_d$, light gray) the \HII Region for the same LOS shown in the middle panel.}
\label{radii_scatter}
\end{figure}

Fig.\ref{map} shows the $x_\nHI$ distribution across the quasar location at
$t=t_Q$, the end of the RT simulation. The \HII region does not exhibit strong deviations from spherical symmetry.
This result is not unexpected: the radiative energy density inside the \HII region during the early phases of
the evolution is so large that clumps possibly responsible for flux anisotropies are completely ionized and
made transparent.  RT effects are instead apparent in the jagged ionization front (IF), causing the size of the
\HII region to fluctuate along different LOS.
We define the radius of the \HII region, $R_d$, along a given LOS as the distance from the quasar at which
$x_\nHI>10^{-3}$, marking the IF.  The RT-induced scatter in the radius of the \HII region is seen in
Fig. \ref{radii_scatter}, via the probability distribution function (PDF) of $R_d$ resulting from a sample
of 1000 LOS piercing the box through the quasar position.  The mean value, $\langle R_d \rangle = 6.29\pm 0.37$ (1-$\sigma$), matches quite well the one derived from eq. \ref{radius}.
In addition, the uncertainty on $R_d$ induced by RT effects is likely smaller than
the experimental error on $z_Q$ determination\footnote{Accurate quasar redshift
determinations are compromised
by the systematic velocity offset between emission lines of highly-ionized
elements (\ie \CIV and \SiIV) and narrow lines probing directly the host
galaxy (\ie \OIII or CO molecular lines). Fan \etal (2006b) quantify an induced
mean error in the measured redshift of $\Delta z\approx 0.02$, corresponding
to a proper distance of $\sim1.2$ Mpc at $z\sim6$.}.

Next, we derived 1000 mock quasar absorption spectra along the same set of LOS
used for $R_d$. The details of the adopted technique  are given in Gallerani
\etal (2006); in brief, each spectrum is characterized by a spectral
resolution ${\cal R}=\lambda/\Delta \lambda\sim 8000$. To enable comparison with data
each spectrum has been smoothed to ${\cal R}=4500$ and Gaussian noise
has been added, yielding a signal-to-noise ratio $S/N=50$ (see \eg Fan \etal 2006b).

From these spectra we aim at deriving the observed \HII region radius, $R_f$.
In general, $R_f \neq R_d$ due to possible effects of the \lya damping wing
absorption arising from \HI located outside the \HII region and to resonant absorption from \HI inside it.
The definition of $R_f$ is somewhat arbitrary, as the transmissivity of the
IGM at $z\approx6$ is a mixture of dark gaps and transmission peaks (Fan \etal
2006a). As a consequence, the edge of the \HII region cannot be simply
identified with the first point at which the transmitted flux drops to zero.
Two different methods have been used so far in the literature:
(i) $R_f$  corresponds to the red side of the GP trough\footnote{As far as this work is concerned, we are taking into account the \lya GP trough. The analysis of the Ly$\beta$ region could provide a larger $R_f$ size. See Section 3 for further
discussion.},
(ii) $R_f$ is identified by the redshift at which the transmitted flux is $> 0.1$, when
the spectrum is rebinned to $\Delta \lambda$=20\AA~ (Fan \etal 2006a).
We have applied both methods to derive $R_f$ from our synthetic spectra
and found only marginal discrepancies.
The $R_f$ PDF obtained from method (i) is shown in Fig. \ref{radii_scatter} (top panel).
From the Figure a large offset between $\langle R_d \rangle=6.29$ Mpc and $\langle R_f
\rangle=4.25$ Mpc is seen: \ie the size of the \HII region extracted from the
spectra is systematically underestimated. We refer to this effect as
{\it apparent shrinking}.
Also shown in Fig.\ref{radii_scatter} (middle panel) are the template and absorbed spectra,
along with the $n_\nHI$ density distribution as a function of observed
wavelength, for a representative LOS.

The total GP optical depth $\tau$, responsible for the apparent shrinking, is the sum of two contributions:
the damping wing absorption $\tau_d$ arising from \HI outside the \HII region\footnote{If the quasar
\HII region is embedded  in a partially neutral IGM, the \HI outside
the \HII region, can produce
significant absorption at wavelengths that in physical space correspond to the ionized region (see \eg Madau \&
Rees, 2000).}, and the resonant one, $\tau_r$, from residual \HI inside it.
A detailed analysis of the mock spectra shows that, for $x_\nHI = 0.1$, the $\tau_r$ contribution to $\tau$ is dominant.
This can be appreciated from the lower panel of Fig.\ref{radii_scatter}, where $\tau_r$ and $\tau_d$ are plotted
separately along the same representative LOS.\footnote{As $\tau_d$ is $\propto x_\nHI$ outside the HII region, it only becomes comparable to $\tau_r$ when $x_\nHI \simgt  0.5$, and overcomes $\tau_r$ in the case of a complete neutral IGM.}
The substantial contribution of resonant absorption results from
the increase of the average $x_\nHI$ with physical distance from the quasar
due to flux geometrical dilution and attenuation. Close to the
edge we find $\tau_r \approx 400$, on average.

The apparent shrinking introduces a mean systematic underestimate of the physical
\HII region size, $R_d$, by $\Delta R = \langle (R_d-R_f)/R_d \rangle = 0.32$. Note that the amplitude
of $\Delta R$ is well above errors induced by RT effects and uncertainties in the quasar parameters.
$\Delta R$ has a considerable dispersion around the mean value above, mostly due to the large fluctuation
of $R_f$ along different LOS (see Fig. \ref{radii_scatter}).
In addition, there is no specific correlation between $R_d$ and $R_f$ along
different LOS.
Both these effects can be understood from the fact that $R_f$ depends on
the \lya optical depth, which in turn is much more
sensitive than the ionizing continuum opacity to tiny fluctuations of $x_\nHI$
inside the bubble. In order to properly predict such
fluctuations, is very important to perform accurate RT calculations.
\begin{figure}
\centerline{\psfig{figure=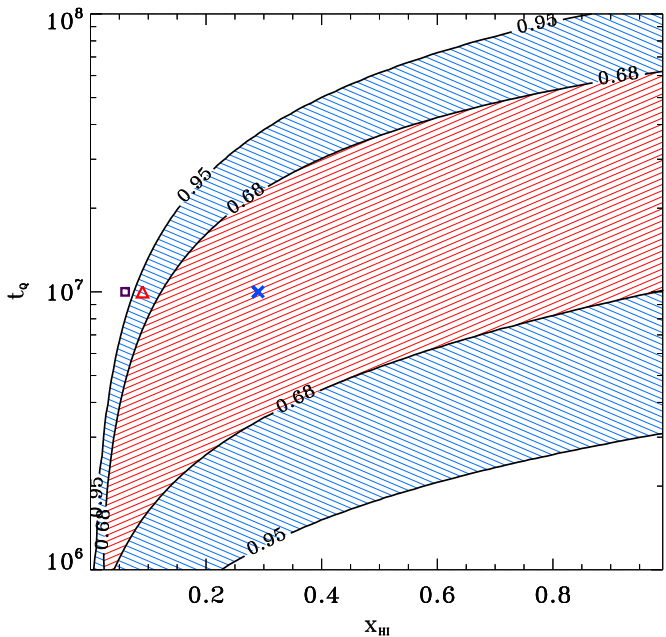,height=8.4cm}}
\caption{Likelihood contours (68 and 95 per cent confident limit) for the $R_f$
distribution in the ($x_\nHI,t_Q$) plane. The cross and triangle
indicate the most likely values $(\hat{x}_\nHI,\hat{t_Q})$
for $R_f$ and $R_d$ distributions, respectively. The square is the
same quantity derived from the observed quasar sample.}
\label{prob_cont}
\end{figure}

In practice, though, the lack of correlation between $R_d$ and $R_f$ makes it very difficult
to precisely derive $R_d$ from the observed spectrum, making it crucial
to quantify the reliability of $x_\nHI$ constraints
obtained from $R_f$ measurements in high-$z$ quasar spectra.  To address this
issue, we have performed a maximum likelihood analysis of our
results. Given the $R_f$ distribution derived from our mock spectra by applying method (ii),
we have calculated its likelihood function (LF) to match a Gaussian distribution,
${\cal G}(R)$, of
mean value given by eq. \ref{radius}, and r.m.s. equal to that of $R_f$. We
impose the following priors:
$\dot{N}_\gamma=5.2\times
10^{56}$ s$^{-1}$, $x_\nHI \in [0,1]$,  $t_Q \in [10^6,10^8]$ yr. For the quasar lifetime $t_Q$ we assume an a lognormal
prior with mean equal to $t_Q=10^7$~yr; uncertainties in $\dot{N}_\gamma$ can be absorbed into $t_Q$ variations.
We have rejected ($x_\nHI$,$t_Q$) pairs yielding $R_d$ values outside the extent of the $R_f$ distribution.
The LF maximum, ($\hat x_\nHI$,$\hat t_Q$), identifies the most likely
values inferred from a given sample of observed quasar spectra.
We find $(\hat{x}_\nHI,\hat{t_Q})=(0.34,10^7 {\rm yr})$, point marked by a cross in Fig.\ref{prob_cont}:
however, these values lay 2-$\sigma$ away from the actual values used in the simulation $(x_\nHI,t_Q)=(0.1,10^7 {\rm yr})$,
\ie those that we would like to recover. As a sanity check we have repeated our maximum likelihood analysis on the
$R_d$ distribution, and indeed we find that the maximum coincides with
the simulation values (triangle in Fig. \ref{prob_cont}). We conclude
that the apparent shrinking effect induces an overestimate of the
$x_\nHI$ by a factor\footnote{This factor could depend slightly on
$x_\nHI$ and, to a lesser extent on $\dot N_\gamma$.} $\approx 3$. It
is worth noting that,
although our $R_f$ distribution has been drawn from a simulation of a single
quasar with fixed intrinsic properties ($\dot{N}_\gamma$, $t_Q$, $z_Q$),
the range of acceptable $x_\nHI$ values is quite large.
For example, fixing $t_Q$ to the simulation value, $10^7$ yr, still allows
$x_\nHI > 0.1,0.07$ to a 1~-~$\sigma$,2-$\sigma$ confidence level. Possible dispersion in the intrinsic properties of the quasar sample
are likely to make the $x_\nHI$ determination even more difficult.

We have also applied the maximum likelihood analysis to a sample of 6
observed QSOs, having $z_Q \in [6.0,6.2]$, whose spectra have been
studied by Fan \etal (2006a). Such authors give the measured radii
scaled to a reference common quasar absolute magnitude
$M_{1450}=-27$. We have then calculated the LF of the sample of
observed spectra to match a Gaussian with mean value given by eq.\ref{radius}
where $\dot{N}_\gamma$ is scaled to
\footnote{We use the scaling: $\dot{N}_\gamma={\dot{N}}_{\gamma,0} \times
10^{(-27-{M}_{1450,0})/2.5}$. We use
${\dot{N}}_{\gamma,0}=(5.2\pm2.5)\times 10^{56}$s$^{-1}$ and
${M}_{1450,0}=-27.2$ which are the estimated values for the
$z=6.28$ QSO SDSS J1030+0524.}
to $M_{1450}=-27$, retaining the observed sample luminosity dispersion.

In this case the LF maximum is $(\hat{x}_\nHI,\hat{t_Q})=(0.06,10^7)$,
shown as a square in Fig.\ref{prob_cont}. By taking into account the overestimate of this result
due to the apparent shrinking effect, and the uncertainties on the location of the maximum induced
by the wide range of acceptable $x_\nHI$ values, we find that our study slightly favors a mostly
ionized universe at $z\approx 6.1$. By varying $\dot{N}_\gamma$ in the range $[2.7,7.7]\times
10^{56}$ suggested by MH04, and scaling the theoretical radii accordingly,
our results are always consistent with $x_\nHI<0.2$.

\section{Summary and Discussion}

In this Letter we have discussed the robustness of constraints on the IGM \HI fraction
inferred from the extent of \HII regions around high redshift quasars by means of their absorption
spectra.

We have performed a combination of state-of-art multiphase
SPH and 3D RT simulations to reliably predict the properties
of a typical high-$z$ quasars \HII region (\eg extent, geometrical shape, inner opacity).
We have found that RT effects do not induce strong deviations from
spherical symmetry. The RT-induced dispersion in the \HII region size
along different LOS is in fact of the order of roughly 6 per cent of the mean radius
which is likely smaller than the typical error induced on $R_d$
estimates by uncertainties in the quasar redshift determinations.

By deriving and analyzing mock spectra through the simulated quasar
environment we have found that the \HII region size deduced from quasar spectra,
$R_f$, typically underestimates the physical one by 30 per cent.

The fact that the observed \HII region sizes can substantially
underestimate the size of the region impacted by the ionizing
radiation of the quasar was already noted by Bolton \& Haehnelt
(2006), but our results gives the first quantitative estimate of the
above underestimation, to which we refer to as apparent shrinking.
It is worthwhile to notice here that the resolution of our simulations
is not sufficient to properly resolve the IGM clumping which could
result in an underestimate of RT effects. To this respect our results
can be considered a lower limit for the amplitude of the apparent
shrinking effect: resolving the high density clumps would result in fact in a higher
inner opacity which is s expected to be the origin of a stronger apparent
shrinking.

The amplitude of the apparent shrinking effect could be decreased by defining
$R_f$ as the red side of the GP trough in the Ly$\beta$ region ($R_f^{\beta}$).
A clear offset between the red side of the GP trough in \lya and
Ly$\beta$ has been observed in the spectrum of the QSO SDSS J1030+0524
(White \etal 2003, MH04. Recently, Bolton \& Haehnelt (2006)
have argued that the ratio $R_f^{\beta}/R_f$ has a well-defined trend with $x_\nHI$,
which could be exploited to constrain such quantity.  However, the
robustness of this result is uncertain given the small number of LOS
used in the study and further investigation is needed.

In addition our analysis shows that, in the case of $x_\nHI=0.1$, the
apparent shrinking is almost completely due to resonant absorption of
residual \HI inside the ionized bubble.
This contribution is highly fluctuating along different LOS, resulting
in a large dispersion of the observed radii distribution, which is
mainly due to the fact that $R_f$ depends on the inner \lya opacity which is highly
sensitive to $n_\nHI$ fluctuations.
This result implies that in order to properly account for
these fluctuations, detailed RT calculations are required; the present work is
unique in this respect, being the only study which accounts for the
detailed effects of 3D RT.

The maximum likelihood analysis we have performed on a sample of 1000
mock spectra shows that the apparent shrinking effect induces an
overestimate of the $x_\nHI$ by a factor $\approx 3$,
if the IGM is only partially ionized ($x_\nHI=0.1$). Moreover, by
applying the above analysis to a sample of observed QSOs, we conclude
that our study favors a mostly ionized universe at $z\sim 6.1$
($x_\nHI\lsim 0.06$), a somewhat different conclusion with respect to
the previous results given in MH04 and  Wyithe \&
Loeb (2004) and Wyithe  \etal (2005), which found a lower limit
for $x_\nHI$ of 0.2 and 0.1 respectively.
Nevertheless, Bolton \& Haenhelt (2006) posted as an outcome of their
analysis that the sizes of the \HII regions in the higher redshift quasar
spectra are consistent with a significantly neutral surrounding IGM,
as well as with a highly ionized IGM.
Furthermore, as mentioned in the Introduction, independent studies
by Yu \& Lu (2005) and Fan \etal (2006a) give evidence
for much lower values of $x_\nHI$.
It is worth noting that our measurement agrees with
the independent determination by Fan \etal (2006) who found
$x_\nHI \simeq 1.3\times 10^{-3}$, based on an analysis of the HII
region size evolution.

Uncertainties remain due to the fact that the range of
acceptable $x_\nHI$ values associated to a given
sample of measured radii is quite large, \eg our $R_f$ sample, having
as most probable value $x_\nHI=0.34$, still allows
$x_\nHI > 0.1,0.07$ to a 1-$\sigma$,2-$\sigma$ confidence level.
This suggests that measurements of the \HII
size in quasar spectra can only provide rough constraints on $x_\nHI$,
as long as the knowledge of intrinsic properties of observed QSOs
remains incomplete.

It is worth noting that quasar spectra could contain additional
useful information on $x_\nHI$. MH04 constrained $x_\nHI$ from an
analysis of the SDSS J1030+0524 absorption spectrum, which is
completely independent from constraing $x_\nHI$ using the radius extent. 
In this case the
$x_\nHI$ determination is based on the difference between the \lya
and Ly$\beta$ optical depths and it is free from the uncertainties
related to the apparent shrinking effect. It is interesting to
notice that, differently from our results, MH04 find that the
slightly smaller size of the \HII region extent inferred from the
\lya and Ly$\beta$ opacity is due to the presence of a damping
wing absorption produced by the \HI in the IGM unperturbed by the
quasar. This difference can be due to the fact that MH04 compute
the neutral hydrogen fraction at the equilibrium with the quasar
ionizing flux which is diluted geometrically, without taking into
account the further attenuation due to the inner opacity. As a
consequence the inner opacity itself is underestimated. We found
instead that the inner opacity, whose correct determination
requires detailed 3D RT calculations, must be taken into account
for a reliable determination of the apparent shrinking effect. A
significant underestimate of the inner (resonant) opacity would
require a higher $x_\nHI$ outside the \HII region in order to
reproduce the observed difference between the apparent shrinking
in the \lya and Ly$\beta$ flux. In any case, though, it is
necessary to await for a larger sample of observed quasar and
simulate a larger parameters space ($x_\nHI$, $t_Q$,
$\dot{N}_\gamma$) before being able to directly compare results
coming from different approaches in the modelling as well as in
the statistical analysis technique adopted.

\section*{Acknowledgments}
We thank B. Ciardi, Z. Haiman and F.S. Kitaura for useful discussions.
AM is supported by the DFG Priority Program 1177.

\label{lastpage}
\end{document}